\documentclass[12pt,epsf]{article}
\usepackage{epsfig,citesort}

\newcommand{\be}{\begin{equation}}
\newcommand{\ee}{\end{equation}}
\newcommand{\bea}{\begin{eqnarray}}
\newcommand{\eea}{\end{eqnarray}}

\def\Frame#1#2{
    \vbox{\hrule
          \hbox
          {\vrule
           \hskip#1
          \vbox{\vskip#1{}
                #2
                \vskip#1}
              \hskip#1
              \vrule}
           \hrule}}

\def\BaseBlock#1#2#3#4#5{%
    \vbox{\setbox0=\hbox{#5}%
        \offinterlineskip
         \hbox{\copy0
         \dimen0=\ht0
         \advance\dimen0by-#1
         \vrule height \dimen0width#2}%
         \hbox{\hskip#3 \dimen0=\wd0
         \advance \dimen0 by -#3
         \advance \dimen0 by #2
         \vrule height #4 width \dimen0}}}

\def\Shadow#1{\BaseBlock{4pt}{2pt}{4pt}{2pt}{#1}}

\begin{document}
\bibliographystyle{unsrt}

\title{The formation of vortex loops (strings)
in continuous phase transitions}
\author{\small \\ Mark J. Bowick$^{(1)}$\hspace{-0.16cm}
\footnote{\tt bowick@physics.syr.edu},\hspace{.3cm}
 Angelo Cacciuto$^{(1)}$\hspace{-.16cm}
\footnote{
\tt cacciuto@physics.syr.edu}\hspace{.2cm} and Alex Travesset$^{(2)}$\hspace{-.16cm}
\footnote{\tt travesse@uiuc.edu}\\
\\\\$^1$ Physics Department, Syracuse University,\\
Syracuse NY 13244-1130, USA \\
$^2$ Loomis Laboratory, University of Illinois at Urbana,\\ Urbana IL
61801, USA\\
}

\date{}

\maketitle

\begin{abstract}
The {\em formation} of vortex loops (global cosmic strings)
in an O(2) linear sigma model in three spatial dimensions
is analyzed numerically. For over-damped Langevin dynamics
we find that defect production is suppressed by an
interaction between correlated domains that reduces the effective
spatial variation of the phase of the order field.
The degree of suppression is sensitive to the quench rate.
A detailed description of the numerical methods used
to analyze the model is also reported.

\end{abstract}

\vfill\newpage

\pagebreak

\section*{Introduction}

The mechanism by which topological defects form in
continuous phase transitions has been a fascinating field
of research for many years with implications for
astroparticle cosmology, particle physics and condensed matter systems
\cite{Riv_rev,Vach_Trieste,Bray_rev,Zurek1,BrDa}.
On the cosmological side it is very likely that the universe
underwent a sequence of symmetry-breaking phase transitions during
its expansion and subsequent cooling after the Big-Bang. If the
disordered phase has symmetry $G$ and the ordered phase a lower
symmetry $H$ then the manifold ${\cal M}$ of possible vacua is
given by the coset space  $G/H$. In many cases the vacuum manifold
${\cal M}$ has some non-trivial topology, allowing for the
appearance of topological defects corresponding to the
non-vanishing homotopy classes of ${\cal M}$. In realistic (finite
quench rate) continuous phase transitions, critical slowing down
then implies that there must be some time during the transition at
which the intrinsic ordering dynamics becomes too sluggish to keep
pace with the quench. It is then inevitable that some neighboring
domains will form with an orientation that produces topological
defects as they coalesce.

On the condensed matter side, where phase transitions occur in
accessible and reproducible laboratory conditions, the production of 
topological defects is familiar and observed in, for example, ferromagnets
\cite{Mermin}, liquid crystals
\cite{Chandrasekhar,DeGennesProst,exp2,exp3,exp1} and superfluids
(both $^4He$ \cite{expMc,exp4,exp5} and $^3He$ \cite{exp6a,exp6}).
\hspace{-0.1cm}\footnote{The most recent improved experiments of the Lancaster
group on the fast adiabatic expansion 
of liquid $^4He$ through the superfluid $\lambda$ transition 
\cite{exp5} do not see any vortex lines, in contrast to
their earlier results \cite{expMc}.}
Most of the work on defects in condensed matter systems has
focused on either the classification of defects
\cite{Kleman,Michel,Bouligand,Mermin}, or the coarsening dynamics
governing the late-time evolution of the defect density
\cite{Bray_rev}. But it is also of considerable interest to
determine the precise mechanism by which defects are produced and
to determine the defect density {\em at} formation
\cite{Zurek1,Kibble1,Kibble2}. This is the problem tackled in this paper
for the case of the linear $O(2)$ sigma model, for which ${\cal
M}$ is the circle $S^1$. We examine numerically the process of
defect production in fixed-rate quenches through the formation,
interaction and coalescence of interacting domains with
well-defined phases and determine the defect density {\em at} the
time of production.

We find there is considerable phase alignment of domains between
their formation and the subsequent production of defects. The
spatial variation of the phase is then smoother than one would
obtain from assuming that domains are statistically independent.
In other words the wandering of the phase on the ground state
manifold $S^1$ from domain to domain is not random {--} domains of
a given phase attract other domains with similar phases. This
clearly reduces the likelihood that domain coalescence will yield
a topological defect.

Although classical and quantum-mechanical continuous phase
transitions have been modelled and systematically investigated,
both analytically
\cite{Kibble1,Kibble2,Zurek1,bray_CM,rivers2,boya_2,boya_NATO,bowick1,rivers3}
and numerically
\cite{Vachaspati,gleiser,calzetta,antunes1,antunes2,gagne}, a
complete understanding is still lacking.

The remainder of the paper consists of five sections.
In section 1 we review mechanisms for the production of
topological defects in continuous phase transitions.
In section 2 we define the problem at hand and outline
our approach.
Section 3 gives the details of the numerical analysis performed
and section 4 gives our results.
Finally we conclude and discuss some possible implications of our work.

\section{The Kibble-Zurek mechanism}

In the Ginzburg-Landau picture continuous phase transitions 
proceed through the growth of 
arbitrarily small amplitude, long-wavelength fluctuations {---}
the so-called spinodal decomposition of the {\em unstable}
symmetric phase to the true ordered phase. 
These spinodal modes grow exponentially in time until cutoff by  
the nonlinearities associated with interactions. When a system undergoing
a continuous phase transition is quenched at a finite
rate from the disordered to the ordered phase, distinct ordered
regions of space ({\bf domains}) will generically lie
at different points on the vacuum manifold. 

In \cite{Kibble1} Kibble proposed a clear mechanism for topological
defect formation in a cosmological context using a simple
ordering-causality argument. The key idea is that spontaneous
symmetry breaking will occur independently in causally
disconnected regions of space. Suppose that order parameter is
uniform within an ordered domain of correlation volume $\xi^3$ and
randomly distributed on the ground-state (vacuum) manifold ${\cal
M}$. Furthermore Kibble assumed that the order parameter between
domains was the smoothest interpolation possible (the {\em
geodesic rule}). In this case one can, in principle, compute the
probability that the coalescence of say three domains leads to a
topologically non-trivial configuration. This probability is a
number of order one so that, roughly speaking, one defect is
formed per correlation volume.

To predict the density of defects at formation in this picture one
must have a theory of the relevant correlation length $\xi$ at
formation. One simple proposal \cite{Kibble1} is that $\xi$
corresponds to the correlation length at the Ginzburg temperature
($T_G$) when ordered domains are stable to thermal fluctuations.
Given the free energy barrier $\Delta F$ (see Fig.~1) between the
true ground state and the unstable high-temperature phase for $T <
T_c$, we have \be k_B T_G\simeq\xi^3(T_G)\Delta F(T_G). \ee

\begin{figure}[h]
\epsfxsize= 3.2in\centerline{\epsfbox{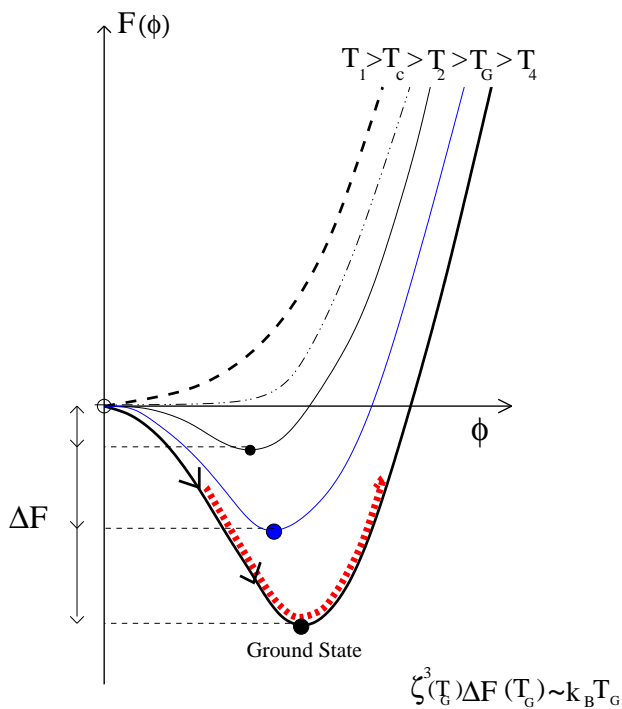}}
\caption{Schematic representation of the free energy curve showing
the disordered ($T>T_c$) and ordered ($T<T_c$) phases for a scalar
field $\phi$ in a $\lambda\phi^4$ theory. The highlighted dashed 
section indicates typical thermal fluctuations for 
temperatures below the Ginzburg
temperature $T<T_G$.} \label{potentiale}
\end{figure}

Above $T_G$  the system can locally jump back and forth between
the high and low temperature phases.
Below $T_G$ this process is thermally suppressed.

This argument ignores the dynamical aspects of the phase
transition and is likely to be inaccurate if defects form at
relatively high temperatures. An alternative {\em non-equilibrium}
approach was proposed by Kibble in a later paper \cite{Kibble2}
and elaborated by Zurek \cite{Zurek1}. Consider a continuous phase
transition proceeding with a finite quench rate. The quench may be
an externally imposed temperature or pressure quench or result
from the expansion of the universe in the cosmological setting. In
the ordered phase a given region of space may attain the ground
state as long as the microscopic dynamics enables it to relax more
rapidly than the quench rate. But in a continuous transition
critical slowing down implies that the intrinsic relaxation rate
becomes arbitrarily slow near the critical temperature. Thus there
is a characteristic time or temperature at which the system cannot
order sufficiently rapidly. The correlation length at this time
provides an estimate of the maximum domain size giving rise to
topological defects.  In causal language typical domain sizes
cannot grow faster than the speed of light and therefore can never
attain the infinite correlation lengths associated with the
critical point.

\begin{figure}[h]
\epsfxsize= 4in\centerline{\epsfbox{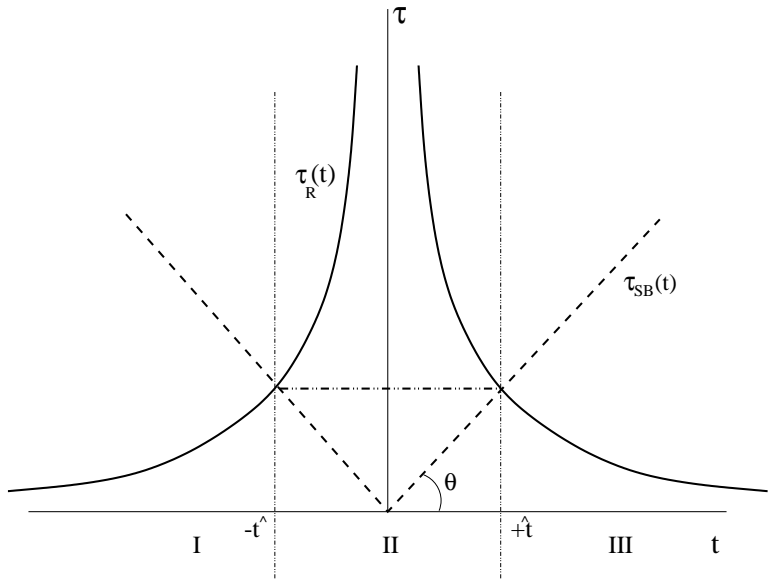}}
\caption{Schematic representation of the relaxation time
$\tau_R(t)$ versus the symmetry breaking time $\tau_{SB}(t)$ of a
system being quenched through a continuous phase transition at a
rate $1/\tau_Q$.} \label{zurek_pic}
\end{figure}

This picture is illustrated in Fig. \ref{zurek_pic}. The two time
scales in the problem are the relaxation time $\tau_R(t)$ and the
symmetry breaking time $\tau_{SB}(t)$  (here taken to be linear).
The dynamics may be divided in three stages. For $t\ll-\hat{t}$
(stage I) the relaxation time of the system is smaller than
$\tau_{SB}(t)$. The field can dynamically relax to equilibrium
while the temperature is falling. When $t\sim-\hat{t}$, the
symmetry breaking time and the relaxation time  become comparable
and when $-\hat{t}<t<\hat{t}$, the situation is reversed. During
stage II $\tau_{SB}(t)\ll\tau_R(t)$, and the time needed by the
system to relax to equilibrium is much larger than the symmetry
breaking time. The system cannot relax and the correlation length
$\xi$ cannot grow as the critical point is approached. Its value
is frozen until, for $t>\hat{t}$,  $\tau_{SB}(t)>\tau_R(t)$ (stage
III) where $\xi$ eventually decreases as expected in the ordered
phase. The faster the quench rate $\tau_Q$, the smaller the angle
$\theta$ indicating the slope of $\tau_{SB}(t)$. Consequently, the
maximum value reached by $\xi$ will be smaller and smaller,
creating a higher defect density. The slower the quench rate, the
larger $\theta$. $\xi$ is allowed to grow to larger values while
approaching the critical point and the density of defects that
form will be smaller (spatially correlated regions have on average
a larger size). Assuming that the correlation length at formation
is the one relative to $t=\hat{t}$, (we can safely assume that
$\xi(-\hat{t})\simeq\xi(\hat{t})$), it is easy to show that a
power law dependence between the freeze-out  correlation length
$\xi^*$ and the quench rate $\tau_Q$ of the phase transition holds
\be \xi ^{*}\sim (\tau _Q)^{\frac{\nu}{\mu+1}}, \ee where
$\tau_R(t)\sim|\varepsilon|^{-\mu}$, $\xi\sim|\varepsilon|^{-\nu}$
and $\varepsilon=t/\tau_Q$. Extensive analytical and numerical
checks have lent support to the Kibble-Zurek mechanism (see
references in the introduction).

\section{Interacting Ordered Domains}

\label{discussion}
In this paper we will model the ordering
kinetics of a non-conserved order parameter by Langevin
dynamics with a global $O(2)$ $\phi^4$ Ginzburg-Landau free energy functional.
We will not treat gauge theories \cite{AR}.
The order parameter $\bar{\phi}$ is zero in the
high-temperature disordered phase and non-zero in the low-temperature
ordered phase. For a continuous transition
$\bar{\phi}$ switches on continuously.
We model the thermal quench via a linearly time varying 
mass term with slope $\tau_Q$.
\be
F(|\phi|)=\int d^dr \left [ \frac{1}{2} (|\nabla
 \phi|)^2+V(|\phi|)\right ] ,
\ee
with a potential of the form
\be\label{potential}
V(|\phi|)=\frac{1}{2}m^2(t)|\phi|^2+
\frac{1}{4}\lambda|\phi|^4 \hspace{1cm}(\lambda>0)
\ee
For $m^2(t)\geq 0$, $|\phi|=0$
and for $m^2(t)<0$, $|\phi|=\pm m(t)/\lambda^{1/2}$.
In the broken-symmetry phase
\be
\phi=\rho e^{i\theta}
\ee
where $\rho\equiv|\phi|$ and $\theta$ is a phase chosen from the
ground-state manifold $S^1$.

Below the critical point ordering progresses via the formation
of ordered domains within which the phase is constant.
Topological defects form as domains coalesce, as described
in the Introduction.

When identifiable stable domains first form they are widely
separated. Between the time they form and the time they coalesce
to produce defects one may expect considerable interaction to take
place since we are dealing with a non-linear field theory. Indeed
angular gradient terms in the broken symmetry regime of the
free-energy should align phases from one domain to the next. This
will suppress wandering of the phase on the ground-state manifold
and lower the probability of forming defects. This is the effect
we establish and quantify in this paper.

\section{Counting Defects: A numerical approach.}\label{numerics}
In this section we discuss how we count topological defects
numerically. We first define an oriented (ordered) domain on the
lattice as the ensemble of spatially connected sites whose phase
difference satisfies the constraint $|\Delta\theta|
<\theta_{\varepsilon}$, where we have introduced a cutoff angle
$\theta_{\varepsilon}$. The number of domains at a given time is
strongly dependent on the value of $\theta_{\varepsilon}$.
Choosing $\theta_{\varepsilon}$ too small will result in too many
domains that are unstable to thermal fluctuations. Taking
$\theta_{\varepsilon}$ too large, on the other hand, means we can
no longer properly track the spatial variation of the phase. The
best compromise is achieved by choosing the largest possible value
of $\theta_{\varepsilon}$ that preserves the topology of the
system. In other words, we look for the effective domains whose
coalescence closely matches the distribution of defects obtained
from the {\em full} field. To be specific we take a $Z_n$
discretization of the circle into $n$ slices $\alpha (k)$
($k=0,1,..n-1$) of width $\Delta=2\pi/n$ and coarse grain the
angular part of the field as follows \be
\theta_i=\Theta^{\alpha(k)} \hspace{.8cm} if \hspace{.3cm}
\theta_i\in \alpha(k) \ee where $i$ is the lattice site and \be
\Theta^{\alpha(k)}=\frac{1}{2}\left (k\Delta+(k+1)\Delta \right )
\hspace{.8cm} k=0,1,..n-1. \ee We then compute the number of
defects using this coarse grained field configuration and compare
it to that obtained using continuous angles. We find that the
largest $\Delta$ that preserves the topology is $\pi/4$,
corresponding to a $Z_8$ discretization. The error in the defect
count resulting from this discretization is on average smaller
than 5\%.

We also address the stability of domains to thermal fluctuations.
For this purpose we introduce a minimum domain size (cut-off) $\Lambda_b$.
The system is kept in contact with a heat bath at
constant temperature $T$ ($T \ll T_c$) throughout the simulation.
$\Lambda_b$ is chosen to be the largest spatially connected domain
generated by thermal fluctuations in the disordered phase.
We now have all necessary tools to explore the dynamics of the effective domains.\\
The following schematic illustrates our strategy:

\begin{picture}(290,390)(-100,0)
\put(-20,330){$\Shadow{\Frame{0.1cm}{\hsize=0.5 \hsize \noindent \begin{center}
   Evolve the system through the phase transition \end{center}}}$}
\put(-20,285){
$\Shadow{\Frame{0.cm}{\hsize=0.5 \hsize \noindent \begin{center}
  Discretize the phase of the field \end{center}}}$}
\put(-40,225){
$\Shadow{\Frame{0.cm}{\hsize=0.6 \hsize \noindent \begin{center}
 Build effective ordered domains out of the discretized field  \end{center}}}$}
\put(-120,120){
$\Shadow{\Frame{0.cm}{\hsize=0.4 \hsize \noindent \begin{center}
Compute the number of defects $N_d$ generated
by the effective domains \end{center}}}$}
\put(90,170){
$\Shadow{\Frame{0.cm}{\hsize=0.4 \hsize \noindent \begin{center}
   Randomize the phase on each effective domain \end{center}}}$}
\put(90,85){
$\Shadow{\Frame{0.1cm}{\hsize=0.4 \hsize \noindent \begin{center}
   Compute number of defects $N_d^{Ran}$ and average over
 30 randomized configurations \end{center}}}$}
\put(-40,20){
$\Shadow{\Frame{0.cm}{\hsize=0.5 \hsize \noindent \begin{center}
  Compare $N_d$ and $N_d^{Ran}$  \end{center}}}$}
\thicklines
\put(85,330){\vector(0,-1){14}}
\put(85,285){\vector(0,-1){14}}
\put(-10,225){\vector(0,-1){45}}
\put(185,225){\vector(0,-1){11}}
\put(185,170){\vector(0,-1){14}}
\put(185,85){\vector(-3,-1){80}}
\put(-10,120){\vector(1,-2){30}}
\end{picture}

\begin{figure}
\epsfxsize= 3.5in\centerline{\epsfbox{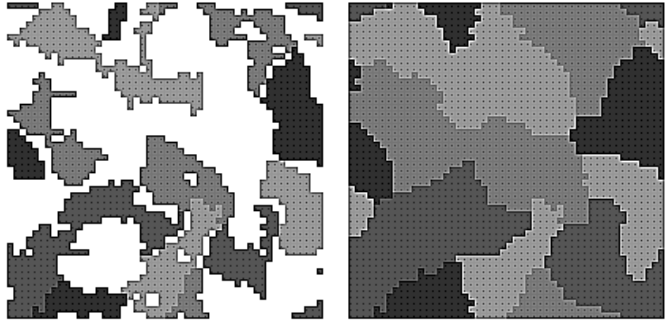}}
\caption{Two dimensional example of the domain reconstruction procedure.
The left image represents a discretized configuration on a 50x50 lattice
(periodic boundary conditions) with a given bubble cut-off $\Lambda_b$, 
and the right image is its layer by layer reconstructed configuration. 
Each color represents a different phase of the field.}
\label{rebuilt}
\end{figure}

The third stage requires more details.
Since we introduced a minimum domain size $\Lambda_b$,
very few domains will form in the early stages
of ordering. Only a small fraction of the lattice is ordered.
One may ask whether these domains have the expected random distribution
of phases. To test this we simultaneously
grow each domain by adding an outer shell of sites of width
equal to one lattice site with the same
field phase. We then recursively grow each domain's external
surface until they
meet in the same region of space and the complete
lattice is filled (see Fig. \ref{rebuilt} for an
illustration of this algorithm applied to a two-dimensional
configuration).
\begin{figure}[t]
\epsfxsize= 4.in\centerline{\epsfbox{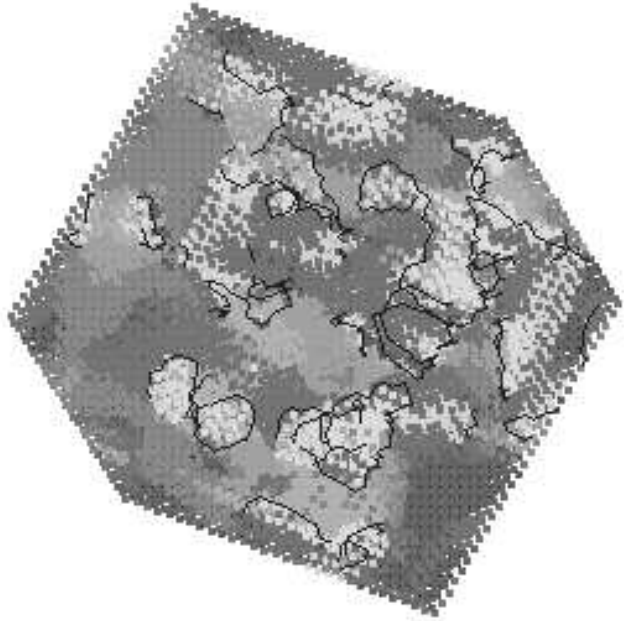}}
\caption{A snap-shot of a typical configuration of the system in a 3D, $L=30$,
lattice with periodic boundary conditions.  Each color represents
a different phase of the field. Dark lines represent the string
defects trapped between the ordered domains.}
\label{snap-shot}
\end{figure}
The final configuration obtained by this construction corresponds
to freely expanding domains with fixed phases. If phases are
randomly distributed across domains the number of defects
determined  before and after randomizing should be statistically
equivalent. As the quench progresses more domains are created over
the lattice. We repeat the same procedure at each time step of the
evolution until almost all the lattice is filled with well-defined
domains. A domain-domain interaction will be traceable by
comparing the number of defects produced. In particular the
difference between the actual number of defects and the number for
the random distribution of phases should grow with time, reaching
a maximum at the time of domain coalescence.

To resolve strings on the lattice
we follow \cite{Vachaspati} and associate a vortex to
each lattice plaquette with a non-trivial phase
winding.
Strings are then constructed by connecting these vortices,
adopting a random reconnection algorithm for the case
of multiple strings passing through the same lattice cell.
To deal with this ambiguity the number of strings at each
time step is obtained by averaging over 15 differently recombined
string configurations \footnote{The uncertainty in this counting
is smaller when the phases are discretized.}.
All simulations were performed on a 300MHz Pentium II
for a total computational time of roughly 800 hours.

\section{Numerical simulation}
We simulated a two-component classical vector field $\vec{\phi}(\vec{r})$
on a three dimensional cubic lattice of side $L=60$ with
periodic boundary conditions.
We evolve the system using a leap-frog numerical implementation of
the Langevin equation in the over-damped regime

\be\label{TDLG}
\frac{1}{\Gamma_0}\frac{\partial \vec{\phi}}{\partial t}=
\vec{\nabla} ^2\vec{\phi}-
\frac{\partial V(\vec{\phi})}{\partial \vec{\phi}}+\vec{\eta}(\vec{r},t)
\ee
where $\Gamma_0$ is a dimensionful constant,  $V(\vec{\phi})$ is given in
Eq.(\ref{potential}) and
$\vec{\eta}(\vec{r},t)$ is Gaussian noise with temperature T
that satisfies the following constraints
\be
\langle\eta_a(\vec{r},t)\rangle=0
\ee
and
\be
\langle\eta_a(\vec{r_1},t_1)\eta_b(\vec{r_2},t_2)\rangle=2T\Gamma_0
\delta_{ab}\delta(\vec{r_1}-\vec{r_2})
\delta(t_1-t_2) .
\ee
It is convenient to rescale to dimensionless variables, to be used 
from now on, as follows
\begin{eqnarray}
x^{'}=m_0 x \hspace{1cm} t^{'}=m_0 t
 && \eta^{'}=\eta\sqrt\lambda/(m_0^3)\nonumber \\
 T^{'}= T\lambda/m_0 &&\phi^{'}=\phi\sqrt\lambda/m_0.
\label{rescaled}
\end{eqnarray}
where $m_0$ is the value of $m(t)$ prior
the quench; when the system is in the symmetric phase.
The rescaled equation, dropping the primes, then becomes
\be
\partial_t\phi=\nabla^2\phi-\alpha(t)\phi-\phi^3+\eta
\label{dyn} \ee where $\alpha(t)=m^2(t)/{m_0}^2$ and we have
chosen $\Gamma_0 m_0=1$. The linear quench is modeled by
\be\label{quench} 
\alpha(t)= \cases{
\hspace{.3cm} 1 \hspace{1.6cm}{\rm for}\hspace{1.15cm} 
t\leq 0\cr
 (1-2\frac{t}{\tau_Q})
\hspace{.4cm} {\rm for}\hspace{.4cm} 
0\leq t\leq\tau_Q
\cr -1\hspace{1.56cm}
{\rm for}\hspace{1.15cm} t\geq\tau_Q} 
\ee 
${\tau_Q}^{-1}$ being the
quench rate. This choice of $\alpha(t)$ enables us to
drive the system from a disordered phase
(convex potential) to an ordered phase 
(sombrero-like potential) in a finite time.
For $t<0$ the system is in thermal equilibrium in the disordered phase.
As $0\leq t\leq \tau_Q$, $\alpha(t)$ linearly decreases until 
it changes sign  passing through the critical point.
When $t\geq \tau_Q$, $\alpha(t)$ stops decreasing 
terminating the quench into the ordered phase.
The limit $\tau_Q\rightarrow 0$ corresponds to an
instantaneous quench while $\tau_Q\rightarrow\infty$ describes an
adiabatic quench. Throughout the simulation the rescaled heat bath
temperature $T$ is held constant. We scanned temperatures ranging
from $T=0.001$ to $T=0.1$. The results reported here are for
$T=0.035$. The equations were numerically solved using $\Delta
x=0.5$ and $\Delta t=0.1$.

In the subsequent analysis we first treat instantaneous quenches
and then compare to the results for a slow quench.

\subsection{Instantaneous Quench: $\tau_Q=0$}
We let the system equilibrate in the disordered phase
($\alpha(t)=1$, $\langle\phi\rangle=0$) in the presence of thermal
noise $\eta$ at $T=0.035$. We sample over 1000 thermalized
configurations to get the minimum ordered domain cut-off
$\Lambda_b$, as previously described. At this temperature and this
lattice size (L=60) we find that the largest thermally generated
bubble in the high temperature phase has a volume of
$V_b\equiv\Lambda_b=79$ spatially connected sites. We then switch
$\alpha(t)=1 \rightarrow \alpha=-1$. This breaks the $O(2)$
symmetry and the system evolves toward its equilibrium value
$\langle\phi\rangle=1$ following the equations of motion defined
in Eq.\ref{dyn}. We calculate the number of defects as explained
in section \ref{numerics}. For each simulation we average the
number of defects over 50 different randomized configurations at
each time step in the dynamics. The number of defects so obtained
is then averaged over a sample of 30 different configurations
obtained from simulations with different random initial
conditions.

\begin{figure}[t]
\epsfxsize= 3in\centerline{\epsfbox{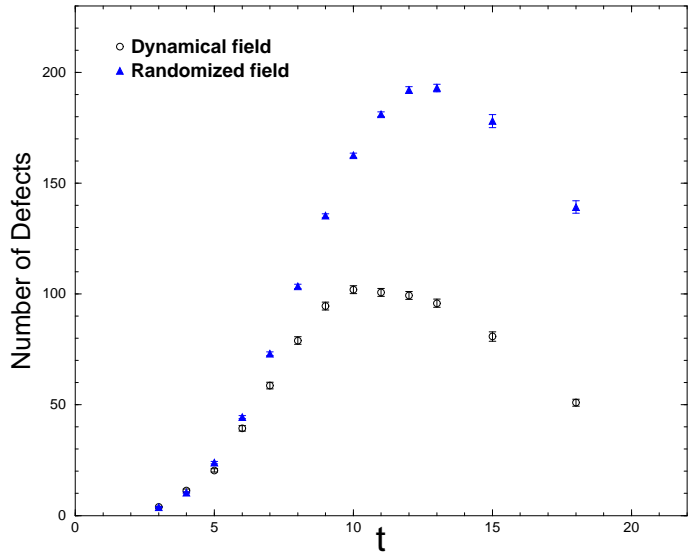}}
\caption{Number of defects generated by the discretized
 field compared to the random-discretized field in an instantaneous quench ($\tau_Q=0$).}
\label{def_tq0}
\end{figure}

Fig. \ref{def_tq0} illustrates  the results of the analysis.
At early times the number of defects produced by both randomized
and non-randomized configurations agrees within statistical
accuracy. This establishes that the order parameter is randomly
varying from one spatial domain to another. At later times the
randomized field configurations produce more and more defects with
respect to its counterpart. This clearly indicates that phase
alignment is occurring as the domains grow. The gap between the
two curves widens until it reaches a maximum at $t=t^*=12\pm 1$.
We associate this particular time with the time of defect
formation. This is supported by the fact that the number of
strings generated by the effective domains reaches its maximum
value at this same time. The number of strings subsequently
decreases as domains coalesce and defects decay. The actual
suppression factor in the number of strings formed is given by
$\gamma=2.0\pm0.2$. We would like to emphasize that this value
should be considered as a lower bound. While counting the number
of strings we scanned our data using different minimum string
length cut-offs $\Lambda_s=4,6,8,10,12,15,18,22$. The data
reported above reflect the analysis  obtained with $\Lambda_s=15$.
Smaller values of $\Lambda_s$ slightly increase the value of
$\gamma$, but also risk over-count defects because many of the
smaller strings could be very short-lived. Larger values of
$\Lambda_s$ do not produce a detectable change in $\gamma$.

\subsection{Slow Quench: $\tau_Q=50$}
\begin{figure}[ht]
\epsfxsize= 3in\centerline{\epsfbox{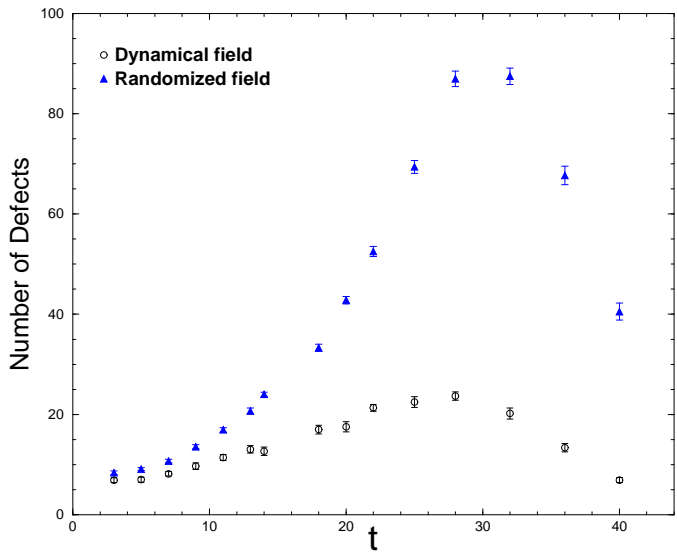}}
\caption{Number of defects generated by the discretized
 field compared to the random-discretized field in a slow quench 
($\tau_Q=50$).}
\label{def_tq50}
\end{figure}

We now consider the effects of a finite-rate quench.
We first equilibrate in the disordered phase ($\alpha(t)=1$,
$\langle\phi\rangle=0$) in the presence of thermal noise
$\eta$ at $T=0.035$ and then initiate the linear quench
according to Eq. \ref{quench}.
The longer time interval over which domains interact, that
between the time they form and the time of defect production,
results in a greater mismatch between the number of defects formed
and the idealized random scenario.
Of course this difference will disappear in the extreme adiabatic limit
as no defects are produced at all in this equilibrium setting.
We expect therefore that there is some finite quench rate which maximizes
the suppression of defect density.
We explored $\tau_Q=50$ as a quench rate fast enough to produce a
large number of defects but still far from the adiabatic limit.
The results are given in Fig. \ref{def_tq50}.
As previously discussed, the gap between the two curves
grows with time until it reaches
a maximum value after which it falls off.
It is revealing to compare
Fig. \ref{def_tq0} and Fig. \ref{def_tq50} to
understand the main physical difference between the
two experiments.
It is obvious that the maximum gap between
the two curves has widened with respect
to the instantaneous quench.
We estimate the formation time
to be $t^*=t=30 \pm 2$ and $\gamma=4.1\pm 0.4$.
The suppression factor $\gamma$ basically doubles while passing from
$\tau_Q=0$ to $\tau_Q=50$.
There is clearly a relationship between $\gamma$ and $\tau_Q$.
As previously discussed we expect that for $\tau_Q\rightarrow\infty$,
$\gamma(\tau_Q)\rightarrow 0$ and would be very interesting to extrapolate
the value of $\tau_Q$ that maximizes $\gamma(\tau_Q)$
but the extremely long computational time required to
achieve this result is prohibitive.

\section{Conclusion and Discussion}
In this paper we have analyzed the effect of a finite quench rate
on the density of topological defects at formation. Detailed
numerical simulations show that the phase angle of ordered domains
aligns in the interval between domain formation and the production
of defects. The effect of this alignment is to reduce the number
of defects that form compared to the simple Kibble mechanism which
assumes that domains are statistically independent at the time of
coalescence. A lower bound on this relative suppression factor is
estimated to be $\gamma=2.0\pm 0.2$ for an instantaneous quench
and $\gamma=4.0\pm 0.4$ for a phase transition with quench time
$\tau_Q=50$. It would be of great interest to systematically vary
$\tau_Q$ to determine the quench rate with optimal defect
suppression. As $\tau_Q$ increases one has to explore larger
correlation lengths which is computationally more demanding. We
hope to undertake this challenge in the near future.

Our results uncover a new underlying feature
of defect formation mechanism that may turn out
to play an important role
in the precise determination of the initial density of defects
generated in a continuous phase transition, a subject of
great interest from both the cosmological and condensed matter
viewpoint. Our study is most directly applicable
to over-damped condensed matter systems
but we think that an analogous effect
exists in the under-damped regime
more appropriate to a relativistic theory.
An analysis in this direction is underway.

\section{Acknowledgements}
The work of MJB, AC and AT was supported by the US Department of
Energy (DOE) under contract No. DE-FG02-85ER40237. The work of AT
was also supported by funding from the materials computation
center, grant NSF-DMR 99-76550 and NSF grant MDMR-0072783.

\end{document}